\documentclass[prd,reprint,showpacs,showkeys]{revtex4-1}
\usepackage{amsfonts}
\usepackage{mdframed}
\usepackage{amssymb}
\usepackage{amsmath}
\usepackage{graphicx}
\usepackage[font={footnotesize,it}]{caption}

\begin{document}

\title{Asymmetric Thin-Shell Wormholes}
\author{S. Danial Forghani}
\email{danial.forghani@emu.edu.tr}
\author{S. Habib Mazharimousavi}
\email{habib.mazhari@emu.edu.tr}
\author{Mustafa Halilsoy}
\email{mustafa.halilsoy@emu.edu.tr}
\date{\today }

\begin{abstract}
Spacetime wormholes in isotropic spacetimes are represented traditionally by
embedding diagrams which were symmetric paraboloids. This mirror symmetry,
however, can be broken by considering different sources on different sides
of the throat. This gives rise to an asymmetric thin-shell wormhole, whose
stability is studied here in the framework of the linear stability analysis.
Having constructed a general formulation, using a variable equation of state
and related junction conditions, the results are tested for some examples of
diverse geometries such as the cosmic string, Schwarzschild, Reissner-Nordstr%
\"{o}m and Minkowski spacetimes. Based on our chosen spacetimes as examples,
our finding suggests that symmetry is an important factor to make a wormhole
more stable. Furthermore, the parameter $\gamma $, which corresponds to the
radius dependency of the pressure on the wormholes's throat, can affect the
stability in a great extent.
\end{abstract}

\maketitle
\affiliation{Department of Physics, Faculty of arts and sciences, Eastern Mediterranean
University, Famagusta, North Cyprus, via Mersin 10, Turkey. }

\section{Introduction}

The history of wormholes goes back to the embedding diagrams of Ludwig Flamm 
\cite{Flamm} in the newly discovered Schwarzschild metric in 1916. Later on,
in 1935, Einstein and Rosen \cite{ER} in search of a geometric model for
elementary particles rediscovered a wormhole as a tunnel connecting two
asymptotically flat spacetimes. The minimum radius of the tunnel, now known
as the throat connecting two geometries, was interpreted as the radius of an
elementary particle. The idea of wormhole did not go in much popularity
until Morris and Thorne \cite{MT} gave a detailed analysis and in certain
sense initiated the modern age of wormholes as tunnels connecting two
spacetimes. It was already stated by Morris and Thorne that the energy
density of such an object, if it ever exists, must be negative; a notorious
concept in the realm of classical physics. In quantum theory, however, rooms
exist to manipulate and live along peacefully with negative energy
densities. Being a classical theory, general relativity must find the remedy
within its classical regime without resorting\ to any quantum. At this
stage, an important contribution came from Visser, who found a way to
confine the negative energy density zone to a very narrow band of spacetime
known as the thin-shell \cite{MV1}. The idea of thin-shell wormholes (TSWs)
became as popular and interesting as the standard wormholes, verified by the
large literature in that context \cite{TSW}. For some more recent works we
refer to \cite{TSW2}. Let us also remark that there have been attempts to
construct TSWs with total positive energy against the negative local energy
density \cite{PE}. This has been possible only by changing the geometrical
structure of the throat, namely from spherical/circular to
non-spherical/non-circular geometry, depending on the dimensionality.
Stability of TSW is another important issue that deserves mentioning and
investigation for the survival of a wormhole \cite{STABILITY}.

\begin{figure}[tbp]
\includegraphics[width=80mm,scale=0.7]{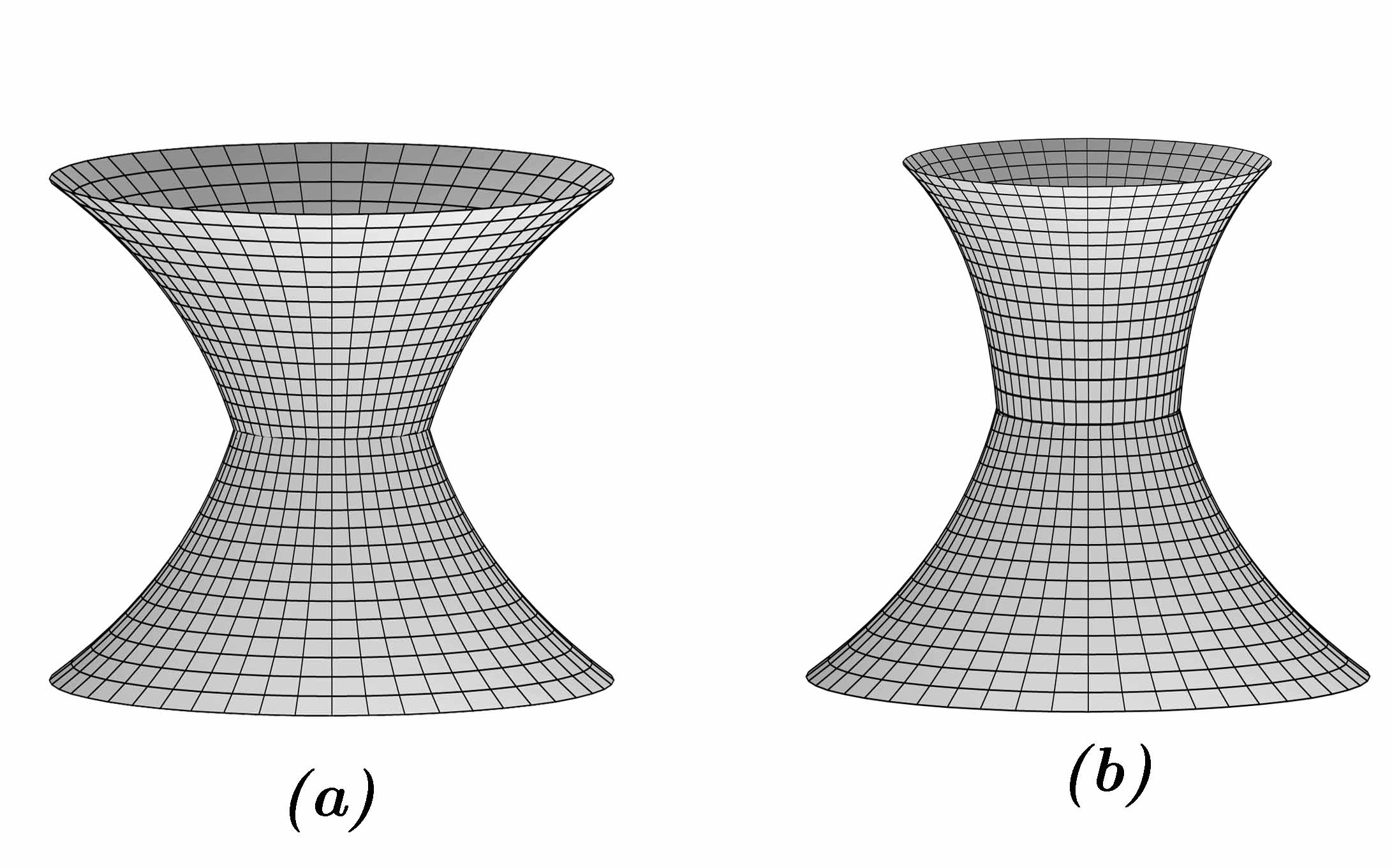}
\caption{Schematic embedding diagram for a symmetric TSW (a) and an ATSW
(b). The throat, on minimum radius hypersurface, separates two
asymptotically flat spacetimes that are not the mirror images of each other.
The upper and lower hyperboloids represent spacetimes with unequal matter
content, joined at the throat in accordance with the junction conditions.
Obviously, the slopes of tangent vectors to the hyperboloids at the throat
are not continuous.}
\end{figure}


In this paper, we introduce TSWs, that are constructed from asymmetric
spacetimes in the bulk \cite{ASYMMETRY}. So far, the two spacetimes on
different sides of the throat, are made from the same bulk material. Our
intention is to consider different spacetimes, or at least different sources
in common types of spacetimes in order to create a difference between the
two sides. Naturally, the reflection symmetry about the throat in the upper
and lower halves will be broken and in consequence new features are expected
to arise which is the basic motivation for the present study. Note that for
non-isotropic bulks, asymmetric TSWs emerge naturally. For example, we
consider Reissner-Nordstr\"{o}m (RN) spacetimes on both sides with different
masses and charges in two sides of the throat; Or two cosmic string (CS)
spacetimes with different deficit angles to be joined at the throat. This
type of TSW, which we dub as asymmetric TSW (ATSW), has not been
investigated so far. For this reason, we will be focusing on such wormholes.
One may anticipate that the asymmetry of the wormhole will have an impact on
particle geodesics, light lensing, and other matters. Asymmetry may act
instrumental in the identification of TSWs in nature, if there exists such
structures. Our next concern will be to study the stability of such ATSW and
novelties that will give rise, if there are any at all. As in the previous
studies, an equation of state (EoS) is introduced at the throat with
pressure and density to be used as the surface energy-momentum tensor. Then,
the Israel junction conditions \cite{Israel} relate these variables within
an energy equation (see Eq. 16), in which $V_{\text{eff}}\left( a\right) $\
is an effective potential. Taking derivative of the energy equation (16)
will naturally yield the equation of motion. Expansion of $V_{\text{eff}%
}\left( a\right) $ about an equilibrium radius of the throat, say $a=a_{0}$,
demands the second derivative $V_{\text{eff}}^{\prime \prime }\
(a_{0})\equiv \left. \frac{d^{2}V_{\text{eff}}\ (a)}{da^{2}}\right\vert
_{a=a_{0}}$\ to be positive. We search for the stability regions of such
ATSW and compare with the symmetric ones.

At this point, we would like to give some information about the stability
analysis and EoS that we shall employ. We adopt a generalized EoS, known as
the variable EoS, defined by $p=p(\sigma ,a)$, where the pressure $p$
depends on both the energy density $\sigma $ and the radius of the shell.
This will bring an extra term defined by $\gamma \equiv -\frac{\partial p}{%
\partial a}\neq 0$, as a new degree of freedom. By this choice, the position
of the shell also will play a vital role in our stability analysis. Since
the expansion of the effective potential about the equilibrium radius will
involve the second derivative of the energy density, emergence of parameter $%
\gamma \neq 0$ in the radial perturbation of the throat will provide us an
extra degree of freedom to achieve $V_{\text{eff}}^{\prime \prime }\
(a_{0})>0$.

Another useful parameter in our analysis will be $\beta ^{2}\equiv -\frac{%
\partial p}{\partial \sigma }$, so that together with $\gamma $, we shall
investigate the stability regions with $V_{\text{eff}}^{\prime \prime }\
(a_{0})>0$. In brief, almost all the detailed information about TSWs studied
so far are also valid for our ATSWs so that we shall refrain from repeating
those arguments. Instead, we shall concentrate on the differences that arise
due to the lack of the mirror symmetry through the throat.

The organization of the paper goes as follows. In section $II$ we give a
general description for TSWs introducing, in particular, the asymmetric
ones. Cosmic string application of ATSW makes the subject matter for section 
$III$. In analogy, the Schwarzschild-Schwarzschild ATSW is considered for
section $IV$. Section $V$ investigates the ATSW constructed from the
Schwarzschild and Reissner-Nordstr\"{o}m spacetimes. Our conclusion to the
paper follows in section $VI$. Some mathematical details for the formalism
are given in the Appendix.

\section{General Formulation}

Having set the general forms of the metrics of the two spacetime geometries
connected by the ATSW%
\begin{equation}
ds_{i}=-f(r_{i})dt_{i}^{2}+\frac{1}{f(r_{i})}dr_{i}^{2}+r_{i}^{2}d\Omega
_{i}^{2};\text{ }i=1,2,
\end{equation}%
we introduce the two spherically symmetric functions $f(r_{i})$ as follows;%
\begin{equation}
f_{i}(r_{i})=k_{i}-\frac{2m_{i}}{r_{i}}+\frac{q_{i}^{2}}{r_{i}^{2}};\text{ }%
i=1,2,
\end{equation}%
where $k_{i}$ are two arbitrary constants expressing the cosmic constants of
the two spacetimes, $m_{i}$ represent the masses and $q_{i}^{2}$ stand for
the sum of the squares of the net electric and magnetic charges of the black
holes in the two spacetimes as observed by a distant observer. Also, in Eq.
(1), $d\Omega _{i}^{2}$\ traditionally stands for $d\theta _{i}^{2}+\sin
^{2}\theta _{i}d\phi _{i}^{2}$. In the case $\left\vert q_{i}\right\vert
\leq m_{i}/\sqrt{k_{i}}$, the singularity is hidden behind an outer horizon
denoted by%
\begin{equation}
r_{+i}=\frac{1}{k_{i}}\left( m_{i}+\sqrt{m_{i}^{2}-k_{i}q_{i}^{2}}\right) ,
\end{equation}%
while for $\left\vert q_{i}\right\vert >m_{i}/\sqrt{k_{i}}$ the spacetime
exhibits a naked singularity.

According to Visser \cite{MV1}, scissoring a region from each spacetime $%
\Sigma _{i}$ with $r_{i}\geq a$, where $a>r_{+i}$, and gluing them together
at their common timelike hypersurface $\partial \Sigma _{i}=\left\{
x|r_{i}-a=0,i=1,2\right\} $ results in a Riemannian, geodesically complete
manifold marked by $\Sigma =\Sigma _{1}\cup \Sigma _{2}$. The hypersurface $%
\Sigma $ which represents a passage between the two spacetimes is a TSW and
we will refer to $r=a$ as the throat of the wormhole. In order to examine
the stability of the throat, one can spot a time-dependent throat,
recognized by $r_{i}=a\left( \tau \right) $, or implicitly%
\begin{equation}
\mathcal{F}_{i}(r_{i},\tau )=r_{i}-a\left( \tau \right) =0;i=1,2,
\end{equation}%
where $\tau $ is the proper time measured by the traveler on the shell.

In the context of Israel junction formalism \cite{Israel}, two conditions
must be satisfied at the wormhole's throat, with the first one expecting a
continuity in the first fundamental form and the second one requiring a
discontinuity for the second fundamental form on the shell. Accordingly, the
first condition gives rise to a unique metric on the shell given by 
\begin{equation}
ds_{shell}^{2}=h_{ab}d\xi ^{a}d\xi ^{b}=-d\tau ^{2}+a^{2}d\Omega ^{2},
\end{equation}%
where now $\theta _{i}=\theta $ and $\phi _{i}=\phi $, while the second
condition demands ($G=c=1$)%
\begin{equation}
\left[ K_{b}^{a}\right] -\delta _{b}^{a}\left[ K\right] =-8\pi S_{b}^{a},
\end{equation}%
where $K_{b}^{a}$ and $K$ indicate the mixed extrinsic curvature tensor and
the total curvature, respectively. Note that $S_{b}^{a}$\ is the
energy-momentum tensor of the shell and $\delta _{b}^{a}$\ is the Kronecker
delta. Furthermore, the square brackets signify a subtraction in the two
sides' curvatures, i.e.%
\begin{equation}
\left[ K_{b}^{a}\right] =K_{b1}^{a}-K_{b2}^{a},
\end{equation}%
with the convention that the indices $a$ and $b$ are those of the shell and
take only\ $\tau $, $\theta $,\ and $\phi $. Having considered this, the
next step should be the calculation of the extrinsic curvature for the two
spacetime geometries. The standard definition of the extrinsic curvature for
each side of the throat is given by (for simplicity, we remove the sub-index 
$i$) 
\begin{equation}
K_{ab}=-n_{\mu }\left( \frac{\partial x^{\mu }}{\partial \xi ^{a}\partial
\xi ^{b}}+\Gamma _{\alpha \beta }^{\mu }\frac{\partial x^{\alpha }}{\partial
\xi ^{a}}\frac{\partial x^{\beta }}{\partial \xi ^{b}}\right) ,
\end{equation}%
where%
\begin{equation}
n_{\mu }=\left( g^{\alpha \beta }\frac{\partial \mathcal{F}}{\partial
x^{\alpha }}\frac{\partial \mathcal{F}}{\partial x^{\beta }}\right) ^{-1/2}%
\frac{\partial \mathcal{F}}{\partial x^{\mu }}
\end{equation}%
is the spacelike four-normal vector satisfying $n_{\mu }n^{\mu }=+1$\ for
the timelike hypersurface, and $\Gamma _{\alpha \beta }^{\mu }$ are the
Christoffel symbols of each bulk geometry. With some algebra, one calculates
the mixed components of the extrinsic curvature tensor as%
\begin{equation}
K_{bi}^{a}=diag\left( \frac{f_{i}^{\prime }+2\ddot{a}}{2\sqrt{f_{i}+\dot{a}%
^{2}}},\frac{\sqrt{f_{i}+\dot{a}^{2}}}{a},\frac{\sqrt{f_{i}+\dot{a}^{2}}}{a}%
\right) ,
\end{equation}%
where a prime $\left( ^{\prime }\right) $\ stands for a total derivative
with respect to the radial distance $r$, and an overdot\ $\dot{a}$\
indicates a total derivative with respect to the proper time $\tau $.
Combining Eqs. (6)\ and (10), together with the energy-momentum tensor of
the shell chosen in the form%
\begin{equation}
S_{b}^{a}=diag(-\sigma ,p,p),
\end{equation}%
turns the second Israel junction conditions\ to the following set of
equations;%
\begin{equation}
\sigma =\frac{-1}{4\pi a}\left( \sqrt{f_{1}+\dot{a}^{2}}+\sqrt{f_{2}+\dot{a}%
^{2}}\right) ,
\end{equation}%
and%
\begin{equation}
p=\frac{1}{8\pi }\left( \frac{f_{1}^{\prime }+2\ddot{a}}{2\sqrt{f_{1}+\dot{a}%
^{2}}}+\frac{f_{2}^{\prime }+2\ddot{a}}{2\sqrt{f_{2}+\dot{a}^{2}}}\right) -%
\frac{\sigma }{2}.
\end{equation}%
Herein, $\sigma $\ is the surface energy density of the shell whereas $p$\
is the angular pressure of the shell. Note that for the matter of symmetry
between the curvature and energy-momentum tensors' elements of $\theta $\
and $\phi $, yields two independent equations instead of three.

By taking derivative of the energy density (12) with respect to the proper
time, one can investigate that the energy conservation relation%
\begin{equation}
\dot{\sigma}=-\frac{2}{a}\dot{a}\left( p+\sigma \right)
\end{equation}%
holds between $\sigma $\ and $p$,\ which alternatively can be expressed as%
\begin{equation}
\sigma ^{\prime }=-\frac{2}{a}\left( p+\sigma \right) .
\end{equation}%
As can be perceived from the latter equation, $p$ and $\sigma $ are not
independent quantities and can be considered related to each other through
an "Equation of State". Although the generic barotropic EoS, $p=p\left(
\sigma \right) $,\ used to be popular in the context of the linearized
stability analysis of wormholes, more recently the variable EoS, $p=p\left(
\sigma ,a\right) $ has been used by Varela \cite{Lobo, Varela} which will be
used in our stability analysis too.

From here on, the method of stability analysis of the wormhole will be quite
similar to that of \cite{Lobo}. With some mathematical manipulations, Eq.
(12)\ can be expressed by%
\begin{equation}
\frac{1}{2}\dot{a}^{2}+V_{\text{eff}}\left( a\right) =0,
\end{equation}%
where one can Taylor expand the effective potential%
\begin{equation}
V_{\text{eff}}\left( a\right) =\frac{1}{2}\left[ \frac{f_{1}+f_{2}}{2}-\frac{%
\left( f_{1}-f_{2}\right) ^{2}}{\left( 8\pi a\sigma \right) ^{2}}-\left(
2\pi a\sigma \right) ^{2}\right]
\end{equation}%
about a presumed equilibrium radius $a_{0}$ to obtain%
\begin{multline}
V_{\text{eff}}\left( a\right) =V_{\text{eff}}\left( a_{0}\right) +V_{\text{%
eff}}^{\prime }\left( a_{0}\right) \left( a-a_{0}\right) + \\
\frac{1}{2}V_{\text{eff}}^{\prime \prime }\left( a_{0}\right) \left(
a-a_{0}\right) ^{2}+O^{3}(a-a_{0}).
\end{multline}

Evidently, the first two terms on the right-hand side of this expansion
become zero; the first as a consequence of Eq. (17)\ and the second, $a_{0}$
represents an equilibrium radius. This implies%
\begin{equation}
V_{\text{eff}}\left( a\right) \simeq \frac{1}{2}V_{\text{eff}}^{\prime
\prime }\left( a_{0}\right) \left( a-a_{0}\right) ^{2}
\end{equation}%
where $V_{\text{eff}}^{\prime \prime }\left( a_{0}\right) $ can be
calculated through calculating consecutive derivations of Eq. (17)\ and
substituting for $\sigma (a_{0})$ and $\sigma ^{\prime }(a_{0})$\ from Eqs.
(12) and (15). However, one must assure that the derivation process is taken
carefully since a second derivative of $\sigma $ arises in $V_{\text{eff}%
}^{\prime \prime }\left( a_{0}\right) $ due to the variable EoS.
Mathematically speaking, this amounts to%
\begin{equation}
\sigma ^{\prime \prime }=\frac{2}{a}\left[ \frac{3}{a}\left( p+\sigma
\right) -p^{\prime }\right] .
\end{equation}%
In the most general case, when the variable EoS $p=p\left( \sigma ,a\right) $
is taken into account, we have%
\begin{equation}
p^{\prime }=\sigma ^{\prime }\frac{\partial p}{\partial \sigma }+\frac{%
\partial p}{\partial a},
\end{equation}%
and%
\begin{equation}
\sigma ^{\prime \prime }\left( a_{0}\right) =\left. \frac{2}{a^{2}}\left[
\left( p+\sigma \right) \left( 3+2\frac{\partial p}{\partial \sigma }\right)
-a\frac{\partial p}{\partial a}\right] \right\vert _{a=a_{0}}.
\end{equation}%
From Eq. (19),\ we are interested in the cases where $V_{\text{eff}}^{\prime
\prime }\left( a_{0}\right) >0$; these are the stable equilibrium states.
Having collected Eq. (2)\ for the radially symmetric functions $f_{1}$ and $%
f_{2}$, together with Eqs. (12), (15) and (20)$\ $for $\sigma $ and its
derivatives,$\ V_{\text{eff}}^{\prime \prime }\left( a_{0}\right) $ can
explicitly be acquired as%
\begin{widetext}
\begin{multline}
V_{\text{eff}}^{\prime \prime }\left( a_{0}\right) =-8\pi \left\{ \left(
4\beta ^{2}+3\right) \sigma _{0}^{2}+2\left( \gamma a_{0}+2p_{0}\beta
^{2}+3p_{0}\right) \sigma _{0}+4p_{0}^{2}\right\} +\frac{\Delta ^{\prime
\prime }}{2}+ \\
\frac{\left[ \left( 4\beta ^{2}-1\right) \sigma _{0}^{2}+\left( \left(
4\beta ^{2}-10\right) p_{0}+2a_{0}\gamma \right) \sigma _{0}-12p_{0}^{2}%
\right] \delta -2a_{0}\sigma _{0}\left( \sigma _{0}+2p_{0}\right) \delta
^{\prime }-\frac{1}{2}a_{0}^{2}\sigma _{0}^{2}\delta ^{\prime \prime }}{%
32\pi ^{2}a_{0}^{4}\sigma _{0}^{4}}
\end{multline}%
\end{widetext}
in which for simplicity we introduced $\delta =\left( f_{20}-f_{10}\right)
^{2}$ and $\Delta =\left( f_{10}+f_{20}\right) .$ Here $f_{i0}$, $\sigma
_{0} $ and $p_{0}$ are the appropriate values for $f_{i}$, $\sigma $ and $p$
at $a_{0}$ given by%
\begin{equation}
f_{i0}=k_{i}-\frac{2m_{i}}{a_{0}}+\frac{q_{i}^{2}}{a_{0}^{2}};i=1,2,
\end{equation}%
\begin{equation}
\sigma _{0}=\frac{-1}{4\pi a_{0}}\left( \sqrt{f_{10}}+\sqrt{f_{20}}\right) ,
\end{equation}%
and%
\begin{equation}
p_{0}=\frac{1}{8\pi }\left( \frac{f_{10}^{\prime }}{2\sqrt{f_{10}}}+\frac{%
f_{20}^{\prime }}{2\sqrt{f_{20}}}\right) -\frac{\sigma _{0}}{2},
\end{equation}%
respectively. Also, $\beta ^{2}$ and $\gamma $ are introduced as the partial
derivatives of $p$ with respect to $\sigma $\ and $-a$, correspondingly;%
\begin{equation}
\beta ^{2}\equiv \frac{\partial p}{\partial \sigma },
\end{equation}%
and%
\begin{equation}
\gamma \equiv -\frac{\partial p}{\partial a},
\end{equation}%
in which $\beta ^{2},\gamma \in 
\mathbb{R}
$.\ 

Hereafter, the potential (23) will be employed in order to analyze the
stability of an ATSW at its throat. During the last 30 years, this has been
done by different authors for diverse spacetimes connected by a symmetric
TSW. For example, a Schwarzschild-Schwarzschild (S-S)\ wormhole was studied
in \cite{MV1} which in the framework of the present article can be evoked by
setting%
\begin{equation}
\left\{ 
\begin{array}{c}
k_{i}=1 \\ 
m_{i}=M \\ 
q_{i}=0 \\ 
\gamma =0%
\end{array}%
\right. .
\end{equation}%
Similarly, the results for $V_{\text{eff}}^{\prime \prime }\left(
a_{0}\right) $ of an RN-RN wormhole, as the one that has been analyzed in 
\cite{RNRN},\ are immediate by setting%
\begin{equation}
\left\{ 
\begin{array}{c}
k_{i}=1 \\ 
m_{i}=M \\ 
q_{i}=Q \\ 
\gamma =0%
\end{array}%
\right. .
\end{equation}

Let us comment that in the stability analysis of thin-shells (not TSWs), the
spacetime geometries of the two sides of the thin-shell are different.
Actually, this is how a physical thin-shell can be defined; roughly
speaking, something whose two sides can be distinguished. Otherwise, our
thin-shell will be merely an imaginary shell in the spacetime.

Now, one may ask the question can we have a TSW connecting two different
geometries? As long as the existing horizons on the two sides (of both
spacetimes) remain behind radius $a$, the answer is yes. As for this, we
would like to have a thorough look at the more exciting cases of
non-identical universes connecting wormholes. In the following, three cases
are studied in detail: An ATSW with two Cosmic String geometries of
different deficit angles are brought together (CS-CS* ATSW); an ATSW
connecting two Schwarzschild universes of different central masses (S-S*
ATSW); and finally an ATSW which provides a bridge between a Schwarzschild
and a Reissner-Nordstr\"{o}m universe (S-RN ATSW).

\section{A CS-CS TSW with Different Deficit Angles}

The two CS universes are characterized by%
\begin{equation}
\left\{ 
\begin{array}{c}
k_{1}=k \\ 
k_{2}=\left( 1+\eta \right) k \\ 
m_{i}=0 \\ 
q_{i}=0%
\end{array}%
\right. ,
\end{equation}%
where $k$ and $\eta $\ are two constants; $k>0$ and $\eta >-1$. Accordingly,
equation (23) for $V_{\text{eff}}^{\prime \prime }\left( a_{0}\right) $\
simplifies to%
\begin{multline}
V_{\text{eff}}^{\prime \prime }\left( x_{0}\right) =\frac{-4\left( \beta
^{2}+\frac{1}{2}\right) \sqrt{1+\eta }}{x_{0}^{2}}+ \\
4\pi \gamma \sqrt{k}\left[ 1+\sqrt{1+\eta }-\frac{\eta ^{2}}{\left( 1+\sqrt{%
1+\eta }\right) ^{3}}\right] ,
\end{multline}%
where $x_{0}=\frac{a_{0}}{\sqrt{k}}$ is the reduced equilibrium radius. As
can be seen easily, in case of a generic barotropic EoS $\left( \gamma
=0\right) $, $V_{\text{eff}}^{\prime \prime }\left( a_{0}\right) $ in Eq.
(32) immediately becomes zero for $\beta ^{2}=-\frac{1}{2}$, which
surprisingly depends neither on $x_{0}$ nor $\eta $. Notice that this rather
general case reduces to a Minkowski-CS (M-CS) ATSW for $k=1$ when $\eta \neq
0$, and for $k=1$\ when $\eta =0$\ simplifies to a Minkowski-Minkowski (M-M)
TSW, whereas still, the conclusion brought after Eq. (32) holds. On the
other hand, in a more general perspective including a variable EoS, one
obtains $\beta ^{2}$ by solving $V_{\text{eff}}^{\prime \prime }\left(
x_{0}\right) =0$, which leads to%
\begin{equation}
\beta ^{2}=\frac{4\pi x_{0}^{2}\sqrt{k}\gamma }{\sqrt{1+\eta }+1}-\frac{1}{2}%
.
\end{equation}%
Due to the restrictions on $k$\ and $\eta $, the coefficient of $\gamma $\
in Eq. (33) is obviously positive definite. This emphasizes that for
negative values of $\gamma $,\ the stability region always shrinks.
Conversely, any positive value for $\gamma $\ results in a stronger
stability, meaning that now there are more values available for $\beta ^{2}$
to occupy in order to have a positive $V_{\text{eff}}^{\prime \prime }$.
Nevertheless, at equilibrium, a positive value for $\gamma $\ determines a
negative value for $\frac{\partial p}{\partial a}$\ by definition, which
physically means that the pressure on the shell alters negatively with a
change in radius. Hence, if presumably, the pressure is not negative itself,
one may come up with the idea that the minus sign has emerged during the
process of derivation. This in turn shows that for example in the case where 
$\gamma $ is proportional to a powered term, i.e. $x_{0}^{n}$, the power $n$
must be negative. Moreover, for certain values of $k$\ and $\eta $,\ the
coefficient of $\gamma $\ behaves quadratically with respect to the reduced
equilibrium radius $x_{0}$, indicating that for a general form of $\gamma
\propto x_{0}^{n}$\ the universal shape of $\beta ^{2}$ against $x_{0}$\ is
predictable.

For the sake of comparison, the associated functions for $\beta ^{2}$\ are
brought in the following for three arbitrary choices of $\gamma $ at
equilibrium. Herein, coefficients are chosen such that they simplify the
form of $\beta ^{2}$\ to the best for further analysis. Having considered
this, we select%
\begin{equation}
\gamma _{a}=\frac{-1}{4\pi \sqrt{k}}\Rightarrow \beta ^{2}=-\frac{\left( 
\sqrt{1+\eta }-1\right) }{\eta }x_{0}^{2}-\frac{1}{2},
\end{equation}%
\begin{equation}
\gamma _{b}=\frac{-1}{4\pi \sqrt{k}x_{0}}\Rightarrow \beta ^{2}=-\frac{%
\left( \sqrt{1+\eta }-1\right) }{\eta }x_{0}-\frac{1}{2},
\end{equation}%
and%
\begin{equation}
\gamma _{c}=\frac{1}{4\pi \sqrt{k}x_{0}^{2}}\Rightarrow \beta ^{2}=\frac{%
\left( \sqrt{1+\eta }-1\right) }{\eta }-\frac{1}{2}.
\end{equation}

\begin{figure}[tbp]
\includegraphics[width=80mm,scale=0.7]{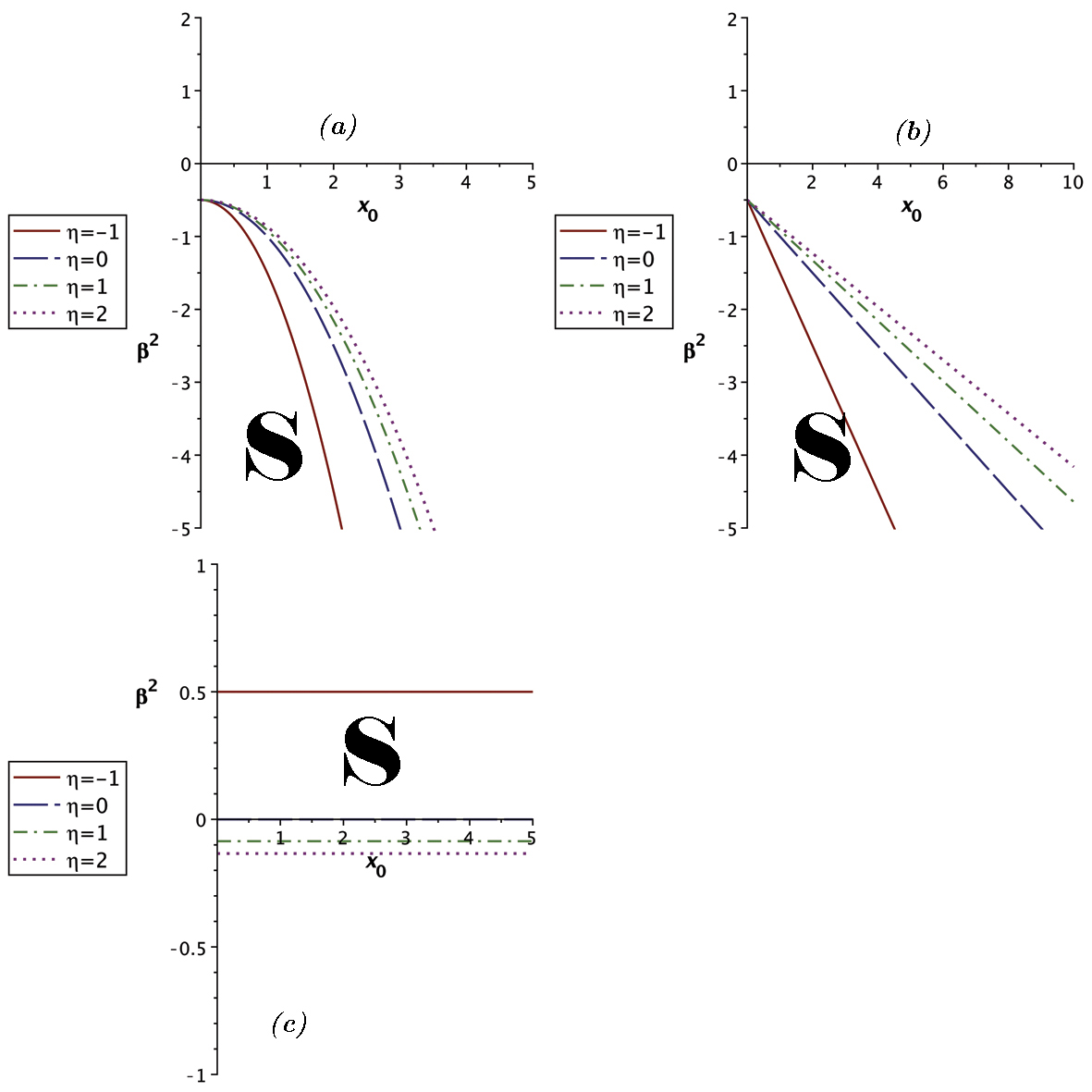}
\caption{{}$\protect\beta ^{2}$ versus $x_{0}$ is plotted for a CS-CS* ATSW
where $\protect\gamma $ is selected as $a)$ $\protect\gamma _{a}=1/4\protect%
\pi \protect\sqrt{k}$, $b)$ $\protect\gamma _{b}=1/4\protect\pi \protect%
\sqrt{k}x_{0}$, and $c)$ $\protect\gamma _{c}=1/4\protect\pi \protect\sqrt{k}%
x_{0}^{2}$. Each figure is drawn for four values of $\protect\eta $. Note
that $\protect\eta =0$ is associated with an M-M TSW. Also, note that $%
\protect\eta $ cannot adopt $-1$\ and the corresponding graphs for $\protect%
\eta =-1$\ are brought as limits. A sign \textbf{S} implies the stable
region. }
\end{figure}


Fig. 2 reflects the graphs for $\beta ^{2}$\ against $x_{0}$ for the three
cases considered above. These figures show different features of stability\
in the vicinity of $a_{0}$ and beyond. The most important aspect is that
there is always a range of values for $\beta ^{2}$\ for which the throat is
stable, apart from any permitted values of $k$ and $\eta $. As another
important outcome, although for $\gamma _{a}$\ and $\gamma _{b}$\ in Eqs.
(34)\ and (35)\ $\beta ^{2}$ is permanently negative in stable states, $%
\gamma _{c}$ in Eq. (36) makes it possible for $\beta ^{2}$ to adopt
positive values when $-1<\eta <0$. This is illustrated in Fig. 3.


\begin{figure}[tbp]
\includegraphics[width=80mm,scale=0.7]{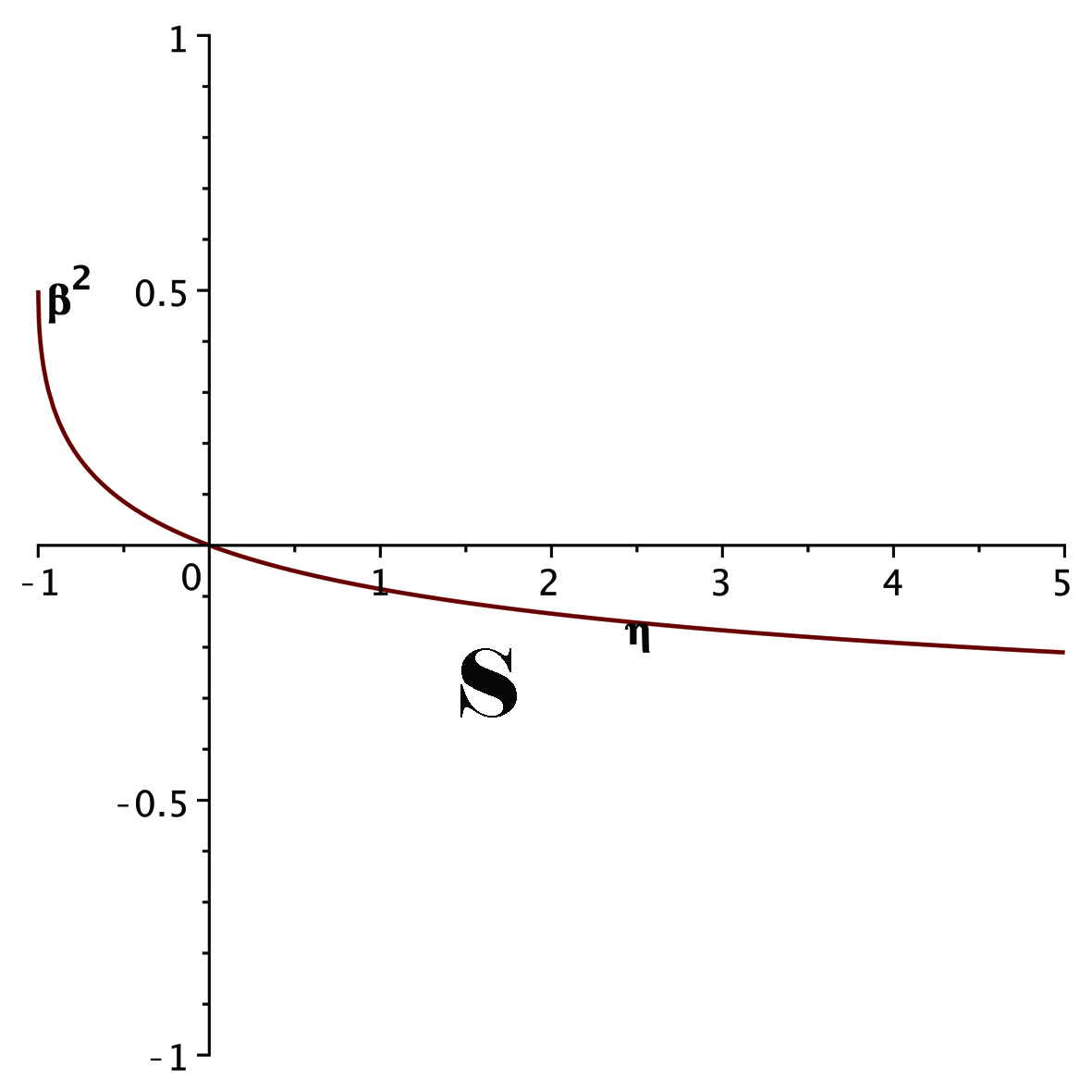}
\caption{The plot of $\protect\beta ^{2}$ against $\protect\eta $ for a
CS-CS ATSW, when $\protect\gamma =\protect\gamma _{c}=1/4\protect\pi \protect%
\sqrt{k}x_{0}^{2}$. The figure implies that for domain $-1<\protect\eta <0$, 
$\protect\beta ^{2}$ can be positive. A sign \textbf{S} implies the stable
region. }
\end{figure}


A slightly different discussion is applied when again a variable EoS is on
the agenda. Solving $V_{\text{eff}}^{\prime \prime }\left( a_{0}\right) =0$
for $\gamma $ leads to%
\begin{equation}
\gamma =\frac{\left( \beta ^{2}+\frac{1}{2}\right) \left( \sqrt{1+\eta }%
+1\right) }{4\pi \sqrt{k}x_{0}^{2}},
\end{equation}%
which can be rewritten by introducing%
\begin{equation}
\gamma ^{\ast }=\frac{4\pi \sqrt{k}\gamma }{\left( \beta ^{2}+\frac{1}{2}%
\right) }
\end{equation}%
in the fashion%
\begin{equation}
\gamma ^{\ast }=\frac{\sqrt{1+\eta }+1}{x_{0}^{2}}.
\end{equation}%
The associated graph indicates that at equilibrium radius, for fixed values
of $k$ and $\beta ^{2}$, $\gamma $ ascends by an increase in $\eta $. The
visualization of $\gamma ^{\ast }$ against $x_{0}$\ is brought in Fig. 4a
for four values of $\eta $ ($\eta =-1$ is brought as a limit).\ Besides,
equation (33) for $\gamma $ exhibits a particular feature that is, for $%
\beta ^{2}=-\frac{1}{2}$, $\gamma $\ vanishes identically. Also, $\gamma
^{\ast }=2$ when $\eta $ is zero; for the case of two Cosmic String (CS)
universes with the same deficit angle. Also, in Fig. 4b the behavior of $%
\gamma ^{\ast }$\ for a constant $x_{0}$\ against parameter $\eta $\ is
projected.


\begin{figure}[tbp]
\includegraphics[width=80mm,scale=0.7]{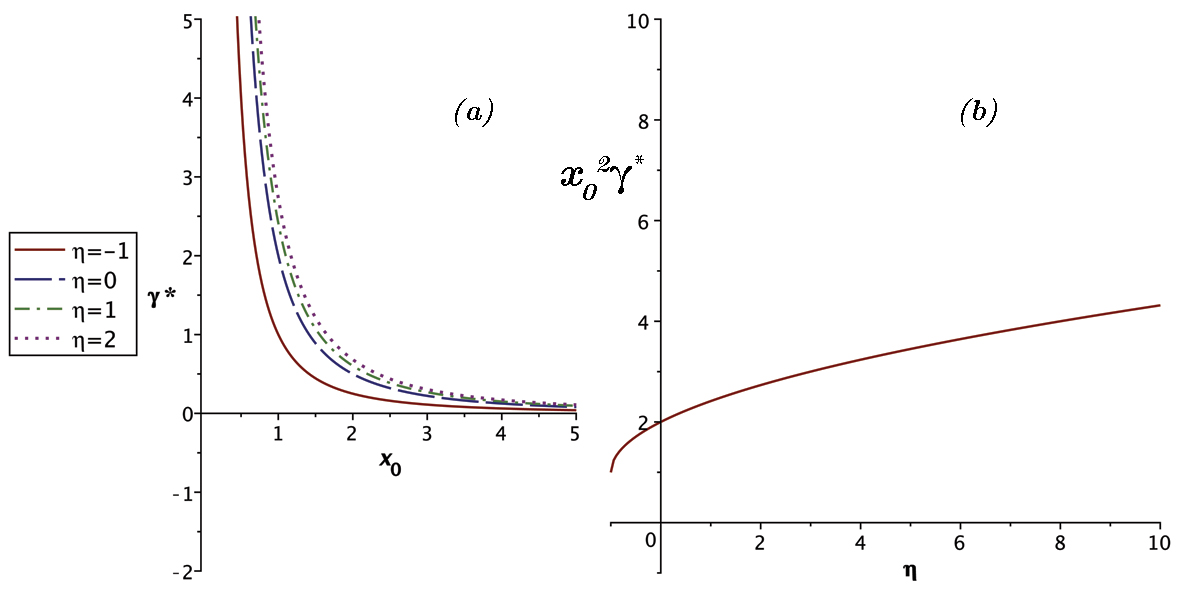}
\caption{For a CS-CS* ATSW in case a variable EoS where $\protect\beta %
^{2}\neq -1/2$, $a)$ shows $\protect\gamma ^{\ast }$ plotted against $x_{0}$
for four different values of $\protect\eta $ ($\protect\eta =-1$ is an
inaccessible\ limit), and $b)$ shows $\protect\gamma ^{\ast }x_{0}^{2}$
plotted against $\protect\eta $. A sign \textbf{S} implies the stable
region. }
\end{figure}


\section{An S-S TSW with Different Central Masses}

As the next example, we look at the case in which the throat provides a
transition between two Schwarzschild geometries possessing different central
masses. In other words, we require%
\begin{equation}
\left\{ 
\begin{array}{c}
k_{i}=1 \\ 
m_{1}=M \\ 
m_{2}=(1+\epsilon )M \\ 
q_{i}=0%
\end{array}%
\right. ,
\end{equation}%
for the two sides' spacetimes, where $\epsilon $\ is a constant; $\epsilon
\geq -1$. Rewriting $V_{\text{eff}}^{\prime \prime }\left( a_{0}\right) $\
from Eq. (23), it can be solved to obtain $\beta ^{2}$ in terms of $\epsilon 
$, $\gamma $\ and $x_{0}$, where $x_{0}\equiv \frac{a_{0}}{M}$\ is the
reduced radius (For the curious reader, the explicit forms of $V_{\text{eff}%
}^{\prime \prime }\left( a_{0}\right) $ and $\beta ^{2}$ are brought in
Appendix A). In the case of a barotropic EoS $\left( \gamma =0\right) $, one
can plot $\beta ^{2}$\ against $x_{0}$\ for various values of $\epsilon $.
It is not hard to see from Eq. (40)\ that for the special case of $\epsilon
=0$, the symmetric S-S wormhole studied by Poisson and Visser in \cite{TSW}
revives; the throat connects two Schwarzschild geometries with the same
central masses.\ Furthermore, it is worth mentioning that for $\epsilon =-1$%
\ the wormhole couples a Schwarzschild spacetime with a flat Minkowski
spacetime. In Fig. 5, the graphs for $\beta ^{2}$ against $x_{0}$ are
depicted for different values of $\epsilon $; $\epsilon =-1.0,-0.5,0.0,0.5,$
and $1.0$. For $\epsilon =0.0$, the shaded areas are the regions of
stability for the wormhole where $V_{\text{eff}}^{\prime \prime }$ is
evidently positive.\ For other values of $\epsilon $ the stable regions are
the same as $\epsilon =0.0$. However, in order to keep the figure less
complicated we did not color those areas. For $\beta ^{2}>0,$ Fig. 5 shows
clearly that deviation from the symmetric TSW with $\epsilon \neq 0.0$ makes
the region of stability smaller. This is the evidence that at least for the
physical meaningful values of $\beta ^{2},$ the more symmetric TSW is, the
more stable it becomes against a radial linear perturbation.

\begin{figure}[tbp]
\includegraphics[width=80mm,scale=0.7]{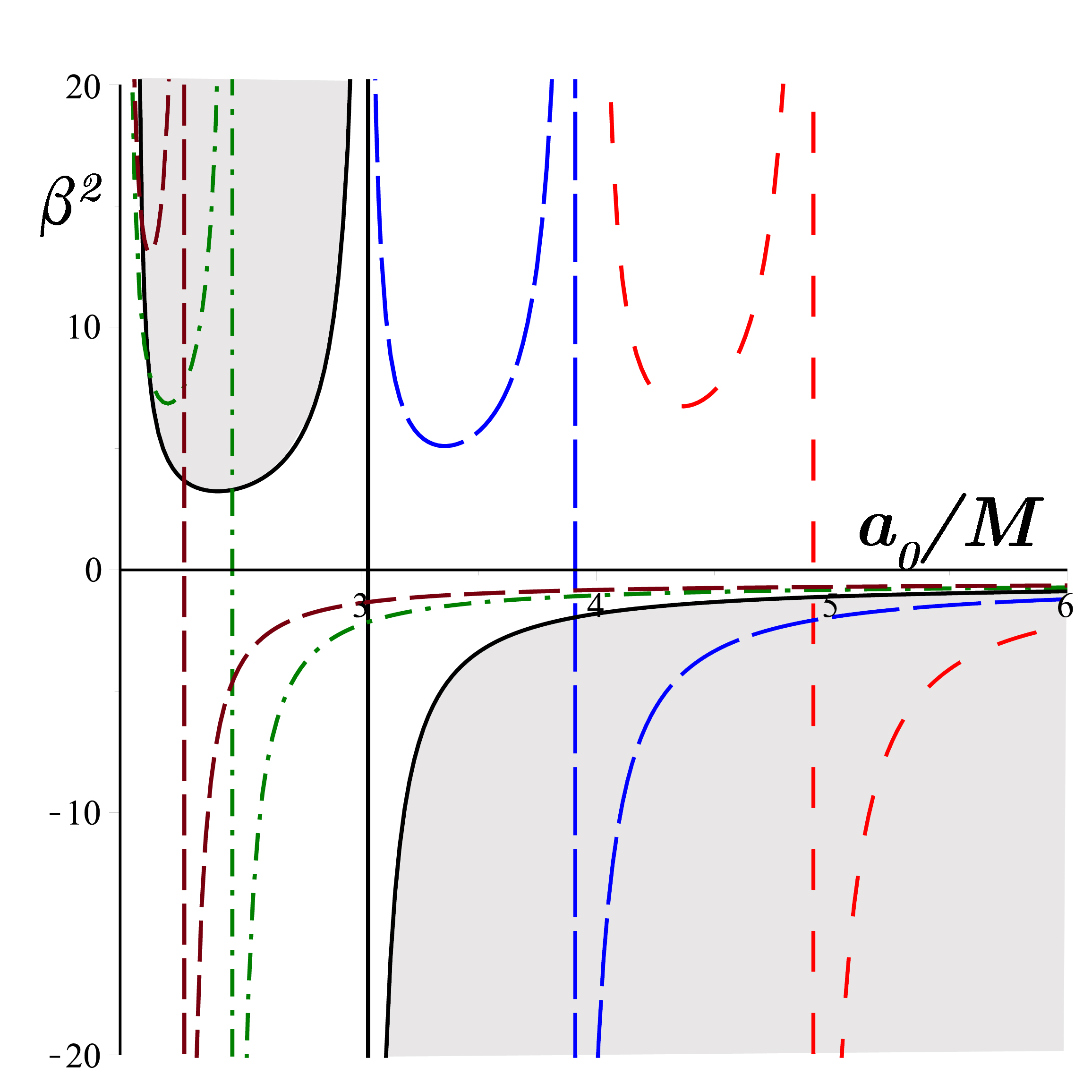}
\caption{{}The plots of $\protect\beta ^{2}$ versus $x_{0}=\frac{a_{0}}{M}$
for an S-S* ATSW with $\protect\gamma =0$ and (from left to right i.e.,
Brown-Dash, Green-Dash dot, Black-Solid, Blue-Long dash and Red-Space dash) $%
\protect\epsilon =-1.0,-0.5,0.0,0.5$ and $1.0.$ The stable regions for $%
\protect\epsilon =0.0$ are shaded. For the other values of $\protect\epsilon 
$ the stable regions are the same as $\protect\epsilon =0.0$ but are not
coloured.}
\end{figure}


As another important example, let us consider the more general EoS $%
p=p\left( \sigma ,a\right) $. Now, $\beta ^{2}$ obtained by setting $V_{%
\text{eff}}^{\prime \prime }\left( a_{0}\right) $ equal to zero will have
terms which include $\gamma $. If one brings these terms together, it can be
apperceived that with a positive slope, $\beta ^{2}$ is linear to $\gamma $.
This implies that the arguments already represented in the previous section
for a CS-CS* ATSW where a variable EoS was discussed, can be summoned here.
Correspondingly, as a general statement concluded from the generic form of $%
\beta ^{2}$ (brought in the Appendix), any negative value given for $\gamma $%
\ results in an instability in the throat while any positive value for $%
\gamma $\ stabilizes ATSW at its equilibrium radius. This is due to the fact
that the coefficient of $\gamma $ in Eq. (A.2) is positive definite.
Likewise, for the general form of $\gamma \propto $ $a^{n}$, if $n>0$/$n<0$,
the growth/decay in stability/instability happens faster with $n$.

Let us\ now choose now a function for $\gamma $\ such that it is
proportional to $a^{-2}$, that is%
\begin{equation}
\gamma =\frac{-1}{\pi M^{2}x_{0}^{2}}
\end{equation}%
merely for the sake of simplification. With the latter choice for\ $\gamma $%
, the behavior of $\beta ^{2}$ versus $x_{0}$ is plotted in Fig. 6 for the
same values of $\epsilon $. We observe that the zoomed-out gesture of the
graphs has altered little from what we had seen in Fig. 5. This is due to
the chosen values of the numerical quantities.

\begin{figure}[tbp]
\includegraphics[width=80mm,scale=0.7]{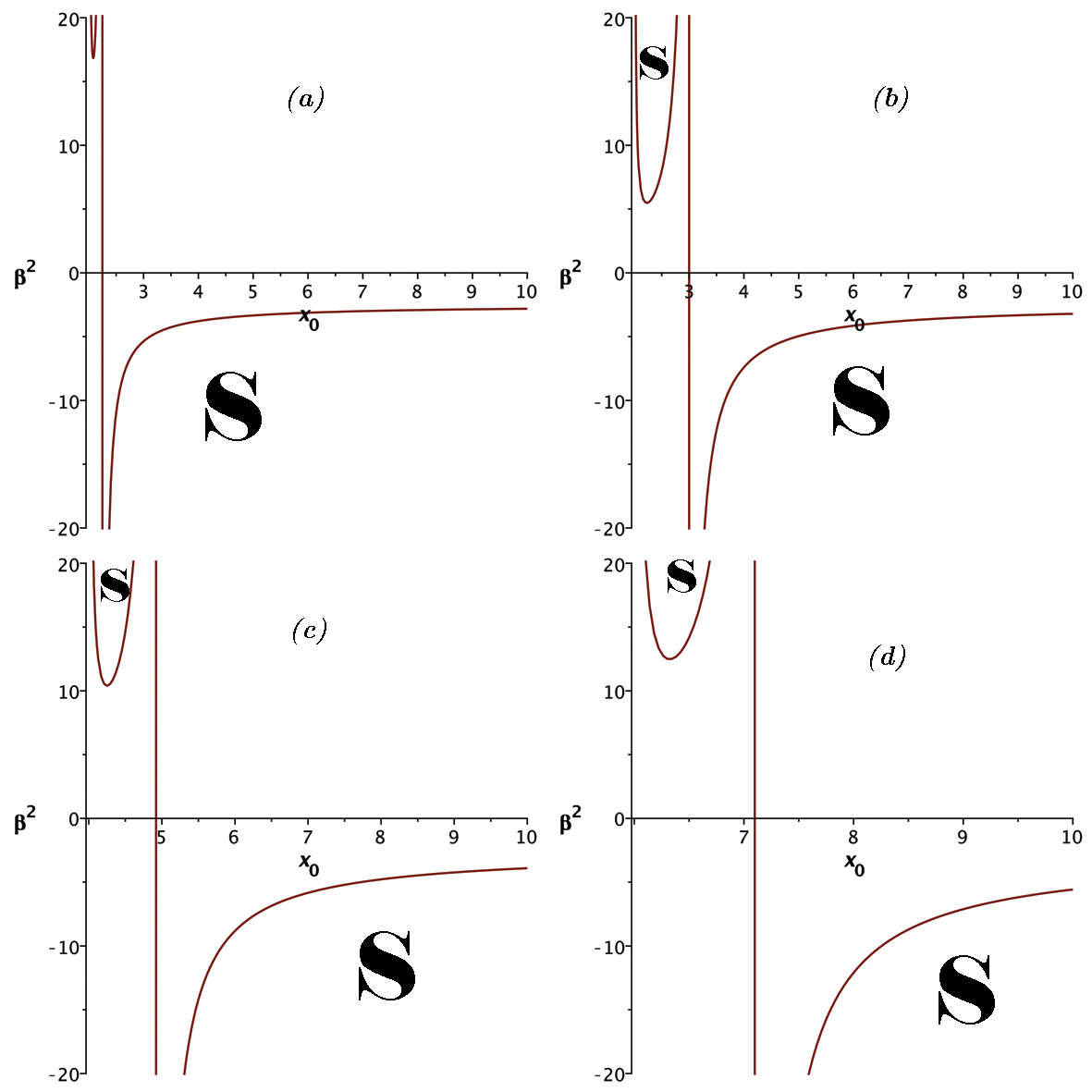}
\caption{{}{}The plots of $\protect\beta ^{2}$ versus $x_{0}$ are given for
an S-S* ATSW when $\protect\gamma =-1/\protect\pi M^{2}x_{0}^{2}$ while $a)$ 
$\protect\epsilon =-1$, $b)$ $\protect\epsilon =0$, $c)$ $\protect\epsilon %
=1 $, and $d)$ $\protect\epsilon =2$. Although the generic shape of the
plots has shown no drastic change compared with the case of a barotropic
EoS, the shifts in the range of $\protect\beta ^{2}$ somehow state that the
stability has decreased. A sign \textbf{S} implies the stable region. }
\end{figure}


On the other hand, things would change significantly if instead of Eq. (41)
we were to pick 
\begin{equation}
\gamma =\frac{1}{\pi M^{2}x_{0}^{2}}.
\end{equation}%
The associated plots for $\epsilon =-1,0,1,2$ are given in Fig. 7, where the
dramatic changes in the regions of stability can be observed. The most
important notion here is that now, for any radial distance, there are always
positive values that $\beta ^{2}$ could adopt.

\begin{figure}[tbp]
\includegraphics[width=80mm,scale=0.7]{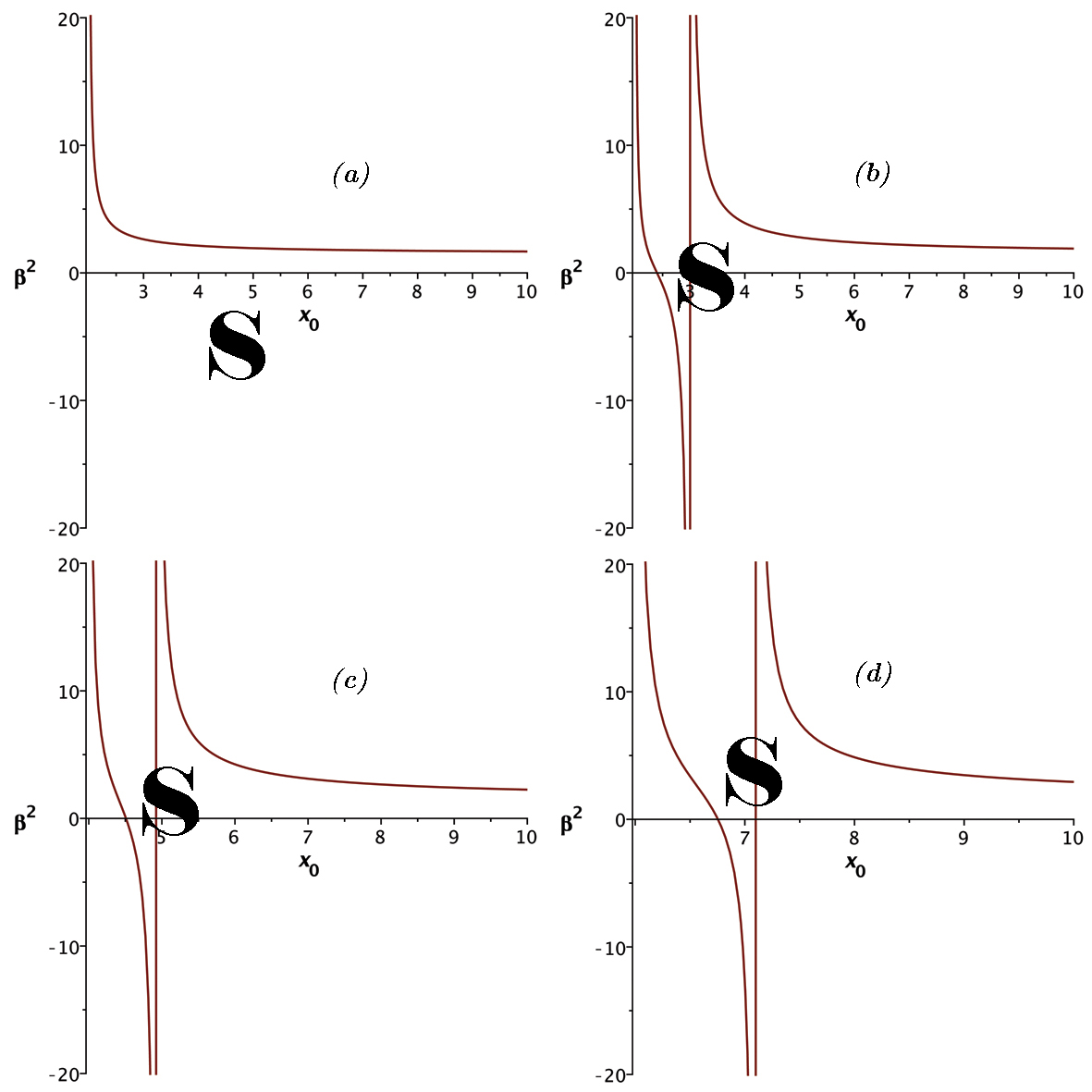}
\caption{{}{}The plots of $\protect\beta ^{2}$ versus $x_{0}$ are given for
an S--S* ATSW when $\protect\gamma =1/\protect\pi M^{2}x_{0}^{2}$ while $a)$ 
$\protect\epsilon =-1$, $b)$ $\protect\epsilon =0$, $c)$ $\protect\epsilon %
=1 $, and $d)$ $\protect\epsilon =2$. Both the general shape and the ranges
of the plot exhibit notable alterations with respect to the case of a
barotropic EoS. A sign \textbf{S} implies the stable region. }
\end{figure}


As another result deduced from Figs. 5-7, although with an increase in $%
\epsilon $\ the areas of stable regions constantly decrease, the region
where $\beta ^{2}$ can possess a positive value increases. The sign of $%
\beta ^{2}$\ is important on account of the association of $\beta $ with the
speed of sound in the material existed on the thin shell; hence a positive
value for $\beta ^{2}$ somehow makes more sense, in physical terms.

As the last example for this section, let us have a look at a rather strange
case where pressure is a function of the radius but not $\sigma $. This
implies that $\beta ^{2}=0$. Therefore, solving $V_{\text{eff}}^{\prime
\prime }=0$ for $\gamma $\ and redefining it as

\begin{equation*}
\gamma ^{\ast \ast }=8\pi M^{2}\gamma ,
\end{equation*}%
\ we arrive at an expression in terms of $x_{0}$ and $\epsilon $ i.e.%
\begin{widetext}
\begin{equation}
\gamma ^{\ast \ast }=\frac{\left[ x_{0}^{2}-3x_{0}\left( 1+\epsilon \right)
+3\left( 1+\epsilon \right) ^{2}\right] \left( x_{0}-2\right) ^{2}\sqrt{%
\frac{x_{0}-2\left( 1+\epsilon \right) }{x_{0}}}+\left[ x_{0}-2\left(
1+\epsilon \right) \right] ^{2}\left( x_{0}^{2}-3x_{0}+3\right) \sqrt{\frac{%
x_{0}-2}{x_{0}}}}{x_{0}^{2}\left( x_{0}-2\right) ^{2}\left[ x_{0}-2\left(
1+\epsilon \right) \right] ^{2}}.
\end{equation}%
\end{widetext}The associated graphs for four values of $\epsilon $\ are
plotted in Fig. 8.

\begin{figure}[tbp]
\includegraphics[width=80mm,scale=0.7]{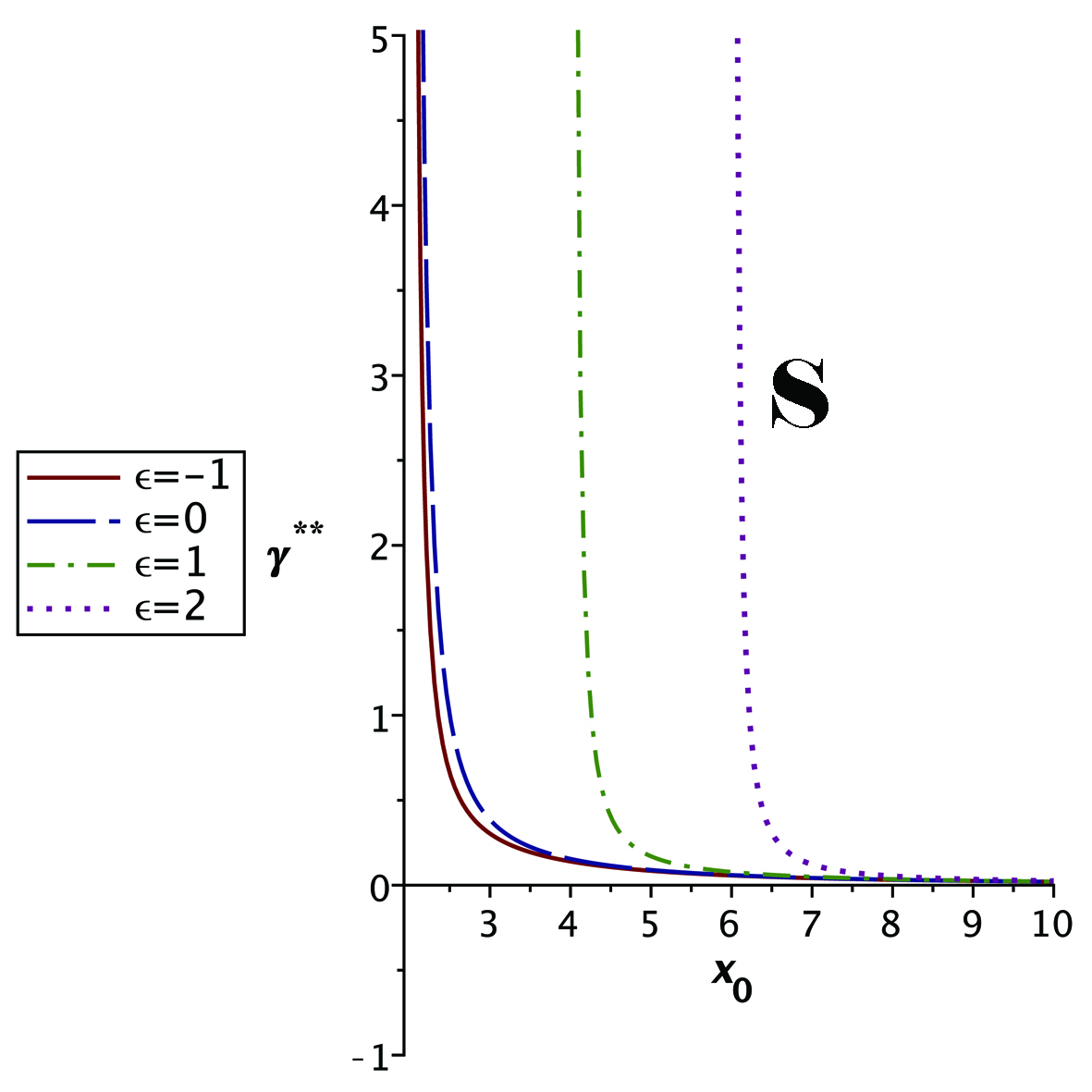}
\caption{$\protect\gamma ^{\ast \ast }$ against $x_{0}$ is plotted for an
S-S* ATSW when $\protect\beta ^{2}=0$, for four diverse values of $\protect%
\epsilon $. Again, $\protect\epsilon =-1$ and $\protect\epsilon =0$ are
related to an S-M ATSW and an S-S TSW, respectively. A sign \textbf{S}
implies the stable region. }
\end{figure}

\begin{figure}[tbp]
\includegraphics[width=80mm,scale=0.7]{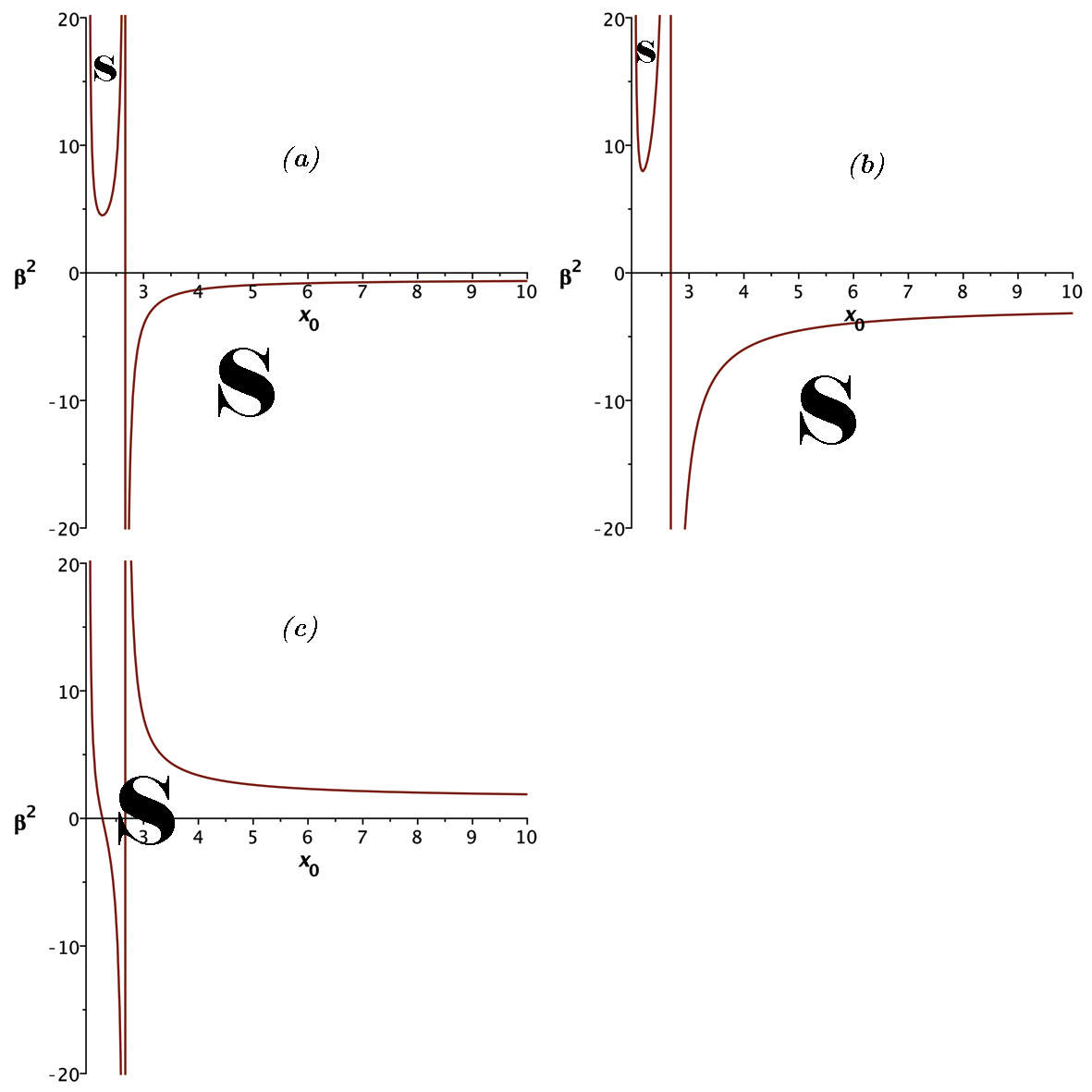}
\caption{{}{}For an S-ERN ATSW, the plot of $\protect\beta ^{2}$ against $%
x_{0}$ are compared for the cases $a)$ $\protect\gamma _{a}=0$, $b)$ $%
\protect\gamma _{b}=-1/\protect\pi M^{2}x_{0}^{2}$, and $c)$ $\protect\gamma %
_{c}=1/\protect\pi M^{2}x_{0}^{2}$. Once more, for a positive $\protect%
\gamma $ the wormhole seems to be more stable than a zero $\protect\gamma $,
which itself is more stable than a negative $\protect\gamma $.}
\end{figure}


\section{An S-RN TSW}

Finally, let us investigate the behavior of an ATSW which connects two
inherently different spacetimes; A Schwarzschild geometry with a central
mass $M$\ and a RN geometry with a central mass $\left( 1+\epsilon \right) M$
and a non-zero total charge $Q$. Hereupon, we demand%
\begin{equation}
\left\{ 
\begin{array}{c}
k_{i}=1 \\ 
m_{1}=M \\ 
m_{2}=(1+\epsilon )M \\ 
q_{1}=0 \\ 
q_{2}=Q%
\end{array}%
\right.
\end{equation}%
for the two sides' spacetimes. Since within the framework of natural units
hired here the mass and the charge have the same dimension of length, we are
allowed to express $Q$ in terms of $M$ in the fashion $Q=\zeta M$, where $%
\zeta $ is a real number. Inserting all these into Eq. (23) results in an
expression for $V_{\text{eff}}$ in terms of $a_{0}$, $\epsilon $, $\zeta $, $%
\beta ^{2}$, $\gamma $ and $M$. Equating $V_{\text{eff}}$ to zero, one can
untangle $\beta ^{2}$ in terms of the remaining parameters. Needless to say,
interplaying with the parameters included\ can produce a huge number of
combinations, each having the potential to be the subject for a separate
detailed study in the future. However, here in this brief account, we wrap
it up with a single example of a very specific case in which $\epsilon =0$
and $\zeta =1$; accordingly, $m_{1}=m_{2}=q_{2}=M$. Evidently, this grants
the special case of Extremal Reissner-Nordstr\"{o}m (ERN) geometry for the
destined spacetime. Hence, in the general case of a radius-dependent
pressure $\left( \gamma \neq 0\right) $, the expression for $\beta ^{2}$\
reduces to%
\begin{multline}
\beta ^{2}=-\frac{3\left( x_{0}-2\right) \sqrt{x_{0}\left( x_{0}-2\right) }%
+x_{0}}{2\left( x_{0}-2\right) \left( 3x_{0}-8\right) }+ \\
\frac{4\pi M^{2}x_{0}^{3}\gamma \left[ \left( -x_{0}+3\right) \sqrt{%
x_{0}\left( x_{0}-2\right) }+\left( x_{0}-2\right) ^{2}\right] }{3x_{0}-8},
\end{multline}%
where in analogy with the previous section, $x_{0}\equiv \frac{a_{0}}{M}$.
Surprisingly, for $\gamma =0$, this $\beta ^{2}$ depicts the same general
configuration as the one in the previous section. This becomes even more
interesting when it is observed that this seemingly similar shape repeats
itself for other permitted values of $\epsilon $\ and $\zeta $ as well. At
this point, the subtle reader would note that $\zeta =0$\ recovers the
discussions represented in the previous section. Fig. 9 illustrates Eq. (43)
for $\beta ^{2}$\ against $x_{0}$\ for three selected functions of $\gamma $%
. These three functions are chosen\ as such, so the deduced diagrams can be
comparable to the ones of the previous section, namely%
\begin{equation*}
\gamma _{a}=0,
\end{equation*}%
\begin{equation*}
\gamma _{b}=\frac{-1}{\pi M^{2}x_{0}^{2}},
\end{equation*}%
and%
\begin{equation*}
\gamma _{c}=\frac{1}{\pi M^{2}x_{0}^{2}}.
\end{equation*}

\section{Conclusion}

In the context of linear stability analysis of TSWs, the wormhole under
study has always been assumed to be symmetric. In this study, however, this
presumption is broken by introducing a new kind of wormholes; ATSWs. To show
that the stability of such peculiar objects can be studied in the context of
linear stability analysis, we have established a general formulation which
was used in the next sections to examine three distinct cases: an ATSW
between two cosmic string universes of different deficit angles, an ATSW
connecting two Schwarzschild geometries of different central masses, and
finally we stepped further to study the stability of an ATSW connecting two
spacetimes of different natures; Schwarzschild and Reissner-Nordstr\"{o}m.
This list can easily be expanded in future works. We have shown that these
objects, with an exotic perfect fluid of EoS $p=p(\sigma ,a)$ on their
surfaces, can be stable. The effective potential of the problem acceptedly
has a more intricate structure compared with the symmetric wormholes. Two
critical parameters, $\beta ^{2}=\frac{\partial p}{\partial \sigma }$ and $%
\gamma =-\frac{\partial p}{\partial a}$, are introduced and analyzed for
each given source in connection with the effective potential. By graphing
their stability diagrams, we qualitatively examined the stability under
various conditions, compared them with each other, and counted some
similarities and differences with the special cases of symmetric TSWs. Most
importantly, it was observed that in case of a barotropic EoS, the stability
diagrams maintain their general shapes, consisting of a bowl-like branch
followed by an asymptotically zero-seeking branch. Comparing the diagrams,
it was stated that in case of barotropic EoS, the thin-shell wormholes
studied here, are most stable at their symmetries. By this,\ it is meant
that for the regions of stability where $\beta ^{2}$ is positive and hence
physically meaningful, the area tends to shrink with any deviation from the
symmetry. Nevertheless, an analytical study can shed more light upon this.
Moreover, in the case of a variable EoS, the general form of the function $%
\gamma $ is proved to be crucial, and can highly manipulate the expected
universal gesture of the stability diagram mentioned above. However, since
choices for the functions of $\gamma $ were up to an arbitrary factor, a
precise numerical and/or analytical assessment is needed to show the full
influence of this pressure's radial-dependency on the stability of ATSWs.


\section{Appendix}

In section $IV$, the effective potential for an S-S* ATSW was discussed
traditionally based on the corresponding graphs of $\beta ^{2}$ against $%
x_{0}$. Nonetheless, the rather unpleasant forms of $V_{\text{eff}}$\ and $%
\beta ^{2}$ are given explicitly here, for the sake of completeness.

The potential is expressed as follows 
\begin{widetext}
\begin{multline}
 V_{\text{eff}}\left( x,\epsilon ,\beta ^{2},\gamma \right) =\frac{1}{%
M^{2}x_{0}^{3}\left( x_{0}-2\right) \left[ x_{0}-2\left( \epsilon +1\right) %
\right] \left[ \sqrt{x_{0}-2}+\sqrt{x_{0}-2\left( \epsilon +1\right) }\right]
^{4}}\times  \tag{A.1} \\
\left\{ x_{0}\sqrt{x_{0}-2\left( \epsilon +1\right) }\left\{ \left[
-16\left( 2\beta ^{2}+1\right) x_{0}^{4}+8\left( 18\beta ^{2}+7\right)
\left( \epsilon +2\right) x_{0}^{3}-4\left[ \left( 46\beta ^{2}+15\right)
\epsilon ^{2}+4\left( 60\beta ^{2}+19\right) \left( \epsilon +1\right) %
\right] x_{0}^{2}\right. \right. \right.  \notag \\
\left. +4\left[ 3\left( 4\beta ^{2}+1\right) \epsilon ^{2}+16\left( 11\beta
^{2}+3\right) \left( \epsilon +1\right) \right] \left( \epsilon +2\right)
x_{0}-24\left( 4\beta ^{2}+1\right) \left( \epsilon ^{2}+8\epsilon +8\right)
\left( \epsilon +1\right) \right] \sqrt{\left( x_{0}-2\right) }  \notag \\
\left. +32\pi \gamma M^{2}\left[ x_{0}-2\left( 1+\epsilon \right) \right]
\left( x_{0}-2\right) ^{2}\left( 2x_{0}-3\epsilon -2\right)
x_{0}^{3/2}\right\}  \notag \\
+32\pi \gamma M^{2}\left[ x_{0}-2\left( 1+\epsilon \right) \right] \left(
x_{0}-2\right) ^{3/2}\left( 2x_{0}-3\epsilon -2\right) ^{2}x_{0}^{5/2} 
\notag \\
-16\left( 2\beta ^{2}+1\right) x^{5}+8\left( 22\beta ^{2}+9\right) \left(
\epsilon +2\right) x_{0}^{4}  \notag \\
-12\left[ \left( 26\beta ^{2}+9\right) \epsilon ^{2}+4\left( 32\beta
^{2}+11\right) \left( \epsilon +1\right) \right] x_{0}^{3}  \notag \\
+4\left[ \left( 44\beta ^{2}+13\right) \epsilon ^{3}+6\left( 84\beta
^{2}+25\right) \epsilon ^{2}+4\left( 3\epsilon +2\right) \left( 104\beta
^{2}+31\right) \right] x_{0}^{2}  \notag \\
\left. -32\left( \epsilon +1\right) \left[ \left( 23\beta ^{2}+6\right)
\epsilon ^{2}+2\left( 56\beta ^{2}+15\right) \left( \epsilon +1\right) %
\right] x_{0}+192\left( 4\beta ^{2}+1\right) \left( \epsilon +1\right)
^{2}\left( \epsilon +2\right) \right\} ;  \notag
\end{multline}%
\end{widetext}while for $\beta ^{2}$\ solving $V_{\text{eff}}=0$ gives rise
to 
\begin{widetext}
\begin{multline}
\beta ^{2}=\frac{1}{2\epsilon \left( x_{0}-2\right) \left[ x_{0}-2\left(
1+\epsilon \right) \right] \left( 4x_{0}^{2}-9x_{0}\epsilon
-18x_{0}+18\epsilon +18\right) }\times   \tag{A.2} \\
\left\{ x_{0}\sqrt{x_{0}-2\left( 1+\epsilon \right) }\left[ -2\sqrt{x_{0}-2}%
\epsilon \left( x_{0}^{2}-\frac{3}{2}\left( \epsilon +2\right) x_{0}+\frac{3%
}{2}\left( \epsilon +1\right) \right) \right. \right.  \\
\left. -8\pi M^{2}\eta x_{0}^{3/2}\left( x_{0}-2\right) ^{2}\left[
x_{0}-2\left( 1+\epsilon \right) \right] \left[ x_{0}-3\left( 1+\epsilon
\right) \right] \right]  \\
+8\pi M^{2}\eta x_{0}^{5/2}\left( x_{0}-2\right) ^{3/2}\left( x_{0}-3\right) 
\left[ x_{0}-2\left( 1+\epsilon \right) \right] ^{2} \\
\left. -2\epsilon \left[ x_{0}^{4}-4\left( 1+\epsilon \right) x_{0}^{3}+%
\frac{3}{2}\left( 3\epsilon ^{2}+17\epsilon +17\right) x_{0}^{2}-18\left(
\epsilon ^{2}+3\epsilon +2\right) x_{0}+18\left( 1+\epsilon \right) ^{2}%
\right] \right\} .  \notag
\end{multline}%
\end{widetext}

\bigskip 

\bigskip 


\begin{thebibliography}{99}
\bibitem{Flamm} L. Flamm, Physikalische Zeitschrift 1\textbf{7}, 448 (1916).

\bibitem{ER} A. Einstein and N. Rosen, Phys. Rev. \textbf{48}, 73 (1935).

\bibitem{MT} M. S. Morris and K. S. Thorne, Am. J. Phys. \textbf{56}, 395
(1988);

M. S. Morris, K. S. Thorne and U. Yurtsever, Phys. Rev. Lett. \textbf{61},
1446 (1988).

\bibitem{MV1} M. Visser, Phys. Rev. D \textbf{39}, 3182 (1989);

M. Visser, Nucl. Phys. H \textbf{328}, 203 (1989).

\bibitem{TSW} P. R. Brady, J. Louko and E. Poisson, Phys. Rev. D \textbf{44}%
, 1891 (1991);

E. Poisson and M. Visser, Phys. Rev. D \textbf{52}, 7318 (1995);

M. Visser, Lorentzian Wormholes from Einstein to Ha-voking (American
Institute of Physics, New York, 1995).

\bibitem{TSW2} E. F. Eiroa and G. F. Aguirre, Phys. Rev. D \textbf{94},
044016 (2016);

E. F. Eiroa and G. F. Aguirre, Eur. Phys. J. C \textbf{76}, 132 (2016);

M. R. Mehdizadeh, M. K. Zangeneh and F. S. N. Lobo, Phys. Rev. D \textbf{92}%
, 044022 (2015);

T. Kokubu and T. Harada, Class. Quant. Grav. \textbf{32,} 205001 (2015);

M. Sharif and M. Azam, Eur. Phys. J. C \textbf{73}, 2407 (2013);

N. M. Garcia, F. S. N. Lobo and M. Visser, Phys. Rev. D \textbf{86}, 044026
(2012);

M. H. Dehghani and M. R. Mehdizadeh, Phys. Rev. D \textbf{85}, 024024 (2012);

X. Yue and S. Gao, Phys. Lett. A \textbf{375}, 2193 (2011);

S. V. Sushkov, Phys. Rev. D \textbf{71}, 043520 (2005);

\bibitem{PE} M. G. Richarte and C. Simeone, Phys. Rev. D \textbf{76}, 087502
(2007); Erratum Phys. Rev. D \textbf{77}, 089903 (2008);

T. Bandyopadhyay and S. Chakraborty, Class. Quantum Grav., \textbf{26},
085005 (2009);

S. H. Mazharimousavi, M. Halilsoy and Z. Amirabi, Class. Quantum Grav., 
\textbf{28}, 025004 (2011);

S. H. Mazharimousavi and M. Halilsoy, Eur. Phys. J. C \textbf{75}, 81 (2015);

S. H. Mazharimousavi and M. Halilsoy, Eur. Phys. J. C \textbf{75}, 271
(2015);

S. H. Mazharimousavi and M. Halilsoy, Eur. Phys. J. C \textbf{75}, 540
(2015);

M. K. Zangeneh, F. S. N. Lobo and M. H. Dehghani, Phys. Rev. D \textbf{92},
124049 (2015).

\bibitem{STABILITY} F. S. N. Lobo, P. Crawford, Class. Quant. Grav. \textbf{%
22}, 4869 (2005);

A. Banerjee, K. Jusufi, S. Bahamonde, Grav. Cosmol. \textbf{24}, 1 (2018);

F. S. N. Lobo, P. Crawford, Class. Quant. Grav. \textbf{21}, 391 (2004);

E. F. Eiroa, G. E. Romero, Gen. Rel. Grav. \textbf{36}, 651 (2004);

E. F. Eiroa, C. Simeone, Phys. Rev. D \textbf{76}, 024021 (2007);

G. A. S. Dias, J. P. S. Lemos, Phys. Rev. D \textbf{82}, 084023 (2010);

E. F. Eiroa, Phys. Rev. D \textbf{78}, 024018 (2008);

J. P. S. Lemos, F. S. N. Lobo, Phys. Rev. D \textbf{78}, 044030 (2008);

F. S. N. Lobo, R. Garattini, JHEP \textbf{1312}, 065 (2013);

E. F. Eiroa, C. Simeone, Phys. Rev. D \textbf{83}, 104009 (2011);

X. Yue, S. Gao, Phys. Lett. A \textbf{375}, 2193 (2011);

S. H. Mazharimousavi, M. Halilsoy, Z. Amirabi, Phys. Rev. D \textbf{89},
084003 (2014).

\bibitem{ASYMMETRY} S. Bahamonde, D. Benisty, E. I. Guendelman,
arXiv:1801.08334;

E. Guendelman, A. Kaganovich, E. Nissimov, S. Pacheva, AIP Conf. Proc. 
\textbf{1243}, 60 (2010);

C. Hoffmann, T. Ioannidou, S. Kahlen, B. Kleihaus, J. Kunz, Phys. Rev. D 
\textbf{95}, 084010 (2017);

\bibitem{Israel} W. Israel, Nuovo Cimento 44B, 1 (1966);

V. de la Cruz and W. Israel, Nuovo Cimento \textbf{51A}, 774 (1967);

J. E. Chase, Nuovo Cimento \textbf{67B}, 136. (1970);

S. K. Blau, E. I. Guendelman and A. H. Guth, Phys. Rev. D \textbf{35}, 1747
(1987);

R. Balbinot and E. Poisson, Phys. Rev. D \textbf{41}, 395 (1990).

\bibitem{Lobo} N. M. Garcia, F. S. N. Lobo and M. Visser, Phys. Rev. D 
\textbf{86}, 044026 (2012);

F. S. N. Lobo, Class. Quant. Grav. \textbf{21}, 4811 (2004).

\bibitem{Varela} V. Varela, Phys. Rev. D \textbf{92}, 044002 (2015).

\bibitem{RNRN} E. F. Eiroa and G. E. Romero, Gen. Rel. Grav. \textbf{36},
651 (2004).
\end{thebibliography}
\end{document}